\documentclass[10pt]{article}
\usepackage{authblk}
\usepackage{graphicx}
\usepackage{wrapfig}
\usepackage{lscape}
\usepackage{rotating}
\usepackage{fullpage}
\usepackage{caption}
\usepackage{float}

\title{Attack Analysis Results for Adversarial Engagement 1 of the DARPA Transparent Computing Program}
\author[2]{Birhanu Eshete}
\author[2]{Rigel Gjomemo}
\author[1]{Md Nahid Hossain}
\author[2]{Sadegh Momeni}
\author[1]{R. Sekar}
\author[1]{Scott Stoller}
\author[2]{V.N. Venkatakrishnan}
\author[1]{Junao Wang}

\affil[1] {Stony Brook University}
\affil [2] {University of Illinois at Chicago}

\date{\today}

\begin{document}
\maketitle

\begin{abstract}
This report presents attack analysis results of the first adversarial engagement event stream for the first engagement of the DARPA TC program conducted in October 2016.
The analysis was performed by Stony Brook University and University of Illinois at Chicago.
The findings in this report are obtained without prior knowledge of the attacks conducted.
\end{abstract}
\thispagestyle{empty}

%\includepdf[pages=-]{uic-engagement1}
%\includepdf[pages=-]{sbu}

% =======UIC report=========

\section{Engagement 1 Analysis Results: UIC}
\subsection{Overview}
We built a tag- and policy-based analysis system for attack detection and forensic analysis. Our system tracks information flows from sources to sensitive sinks in real time, and raises alarms when the policies are violated. In addition, it keeps a \textit{provenance context} with every entity  in the system. Specifically, \textit{provenance context} contains a list of the events and relationships that contribute to the existence of an entity. Using such provenance context, we are able to provide an explanation for the raised alarms under the form of dot graphs. During Engagement 1, we successfully detected the attacks on several TA1 traces and produced graphs with details about those attacks. 

\subsection{Consumed TA1 Data and Challenges}
At UIC, we consumed and analyzed CADETS traces for all three scenarios (BOVIA, PANDEX, and Stretch Period). 

\subsection{Challenges}
We faced several challenges due to data formats and lack of documentation about the traces. The data format challenges were especially frustrating, since they required a large amount of effort to debug and fix. More specifically, we spent around 70\% on this task alone. We report below details about the most prominent issues that we faced.

\subsubsection{Data Format Challenges}
\noindent
\textbf{CADETS}
\begin{itemize}
\item \textit{Missing IPs}. Sometimes, the IP address of a netflow object is not in the definition of the object. Whereas, it is available in EVENT\_Socket or other events. Other times, there is no IP address for a netflow object, neither in the definition nor in other related events. We treated these missing IPs as untrusted and consequently had an increase in the false positives number. 

\item \textit{Inconsistent File Object URLs.} Some versions of file objects have empty URL while other versions have URL. Other times the URL is missing for all the versions of the file objects. In addition, many file paths are in different fields like \texttt{[``properties''][``fdpath'']} or \texttt{[``properties''][``upath1'']} while they are expected to be in URL field of file objects.

\item \textit{Repeated UUIDs}. In some cases, the same UUID is used for different objects like a file and a netflow.

\item \textit{Redundant Data}. There are multiple and unnecessary definitions of the same object.

\item \textit{Unconnected Records}. Some events are expected to be connected to netflow or file objects (like recvmsg), but there are no \textit{SimpleEdge} records for this.

\item \textit{Missing Subject Names}. The process names should be part of their definitions but usually, they can be found in the exec field of other events generated by that process.

\item \textit{Undefined UUIDs}. Some UUIDs are used in events that are not defined in the trace, e.g., sending to a UUID, for which there is not definition.
\end{itemize}

\subsection{Technology Summary}
Our system follows a tag- and policy-based approach for attack detection and forensic analysis. In particular, we use tags to track information flow from predefined sources to sensitive sinks. We use 2 types of tags: 1) integrity tags, and 2) confidentiality tags. Every relevant entity in the system being tracked is associated with such tags and they are propagated to new entities as they are consumed from the TA1 traces. 

\subsubsection{Tags}
\noindent
\textbf{Integrity tags}. Integrity tags represent the degree of \textit{trustworthiness} of the associated system entity. We use \textit{code integrity} tags to represent the trustworthiness of programs and code, and \textit{data integrity} tags to represent the trustworthiness of data. The code integrity tags are as follows: 

\begin{enumerate}
\item \textit{Whitelist}. Used to eliminate background noise. 
\item \textit{Invulnerable}. Used for programs that consume often low integrity data (e.g., browser). They can be downgraded when they execute or load low integrity code.
\item \textit{Benign+authentic}. Used for non malicious programs that are authenticated. These programs can be downgraded when consuming low integrity code or data.
\item \textit{Benign}. Used for non malicious programs, unless exposed to untrusted code or data.
\item \textit{Untrusted}. Used for code from untrusted sources.
\item \textit{Malicious}. Used for programs that exhibit clearly malicious behavior.
\end{enumerate}

The data integrity tags include the last 4 tags of the above list (Benign+authentic, Benign, Untrusted, Malicious).

\noindent
\textbf{Confidentiality tags}.
Confidentiality tags express the degree of secrecy of certain data and the degree of protection they need. They are as follows:
\begin{enumerate}
\item \textit{Public}. Readily available from public sources, no need to protect. Assigned to most code.
\item \textit{Private}. Data with some privacy concerns, but none too specific. Assigned to most data files and data downloaded from most web sites.
\item \textit{Sensitive}. Data with specific security or privacy concerns (e.g., server or host configuration files, emails).
\item \textit{Secret}. Data whose loss can enable impersonation (e.g., \texttt{/etc/shadow}, ssh host or user private keys.
\end{enumerate}

\subsubsection{Detection Policies and Forensics}

Attack detection is guided by several policies, which make use of tags. These policies may deal with integrity (e.g., raise an alarm if a low integrity file is executed), confidentiality (e.g., raise an alarm if secret information flows to low integrity sockets), or a combination of both (e.g., raise an alarm if an untrusted program reads a secret file and sends it to a low integrity socket). 

In addition to the detection policies, our system  contains a list of tag propagation policies, or rules, that govern the propagation of tags through the system entities. These rules are specific to each event and are executed as the events are consumed from the TA1 traces. 

We have implemented several types of policies, that deal mainly with information flow from secret and sensitive data to untrusted sockets and programs. Our policies make use of an initial of sources (list of IPs that are untrusted, a list of files that are secret) and sinks (list of trusted IP addresses, secret and sensitve files). As the records are consumed we maintain a data structure that represents the system entities and their tags and use it to enforce the policies. 

To support forensics, in addition to the tags, we associate a \textit{provenance context} to each system entity. This context contains all the events and entities that, starting from the sources, contribute to the state of a system entity. This provenance context is \textit{propagated forward} and augmented, as new events are consumed from the TA1 traces. For instance, if a browser forks a shell, the fork event is added to the provenance context of that shell. If that shell writes to a file, the context of the shell is copied to the context of the file, and the write event is added to the context of the file. Therefore, the context of the file will contain both the fork event and the write event. If that file is executed and a new process is created, its context is copied to the context of the new process, and the execute event is added to it. 

Using forward propagation of the \textit{provenance context}, we are able to immediately obtain the history of events that contributed to an alarm. In particular, for every alarm that is raised, our system processes the provenance context of the entities involved in the alarm and produces a dot file that represents the history of the alarm. 

\subsection{Results}
In this section, we present our results on the CADETS traces. We successfully detected and reconstructed the history of the attacks on the CADETS machines. For each scenario, we attach the graph of the attack. In these graphs, we depict processes by ovals, file objects by rectangles, and sockets by diamonds. We use labeled and directed edges to represent system calls. The direction of the edges represents the direction of information flow, while the number of the edge labels represents the order of the system calls. We note, that for space purposes and readability, we have removed several edges and nodes from the final result. These represent activities that occur at the same time as the attack but are most likely not connected to it.

\subsubsection{Bovia Scenario}
\begin{sidewaysfigure}[t]
    \includegraphics[width=0.95\textwidth]{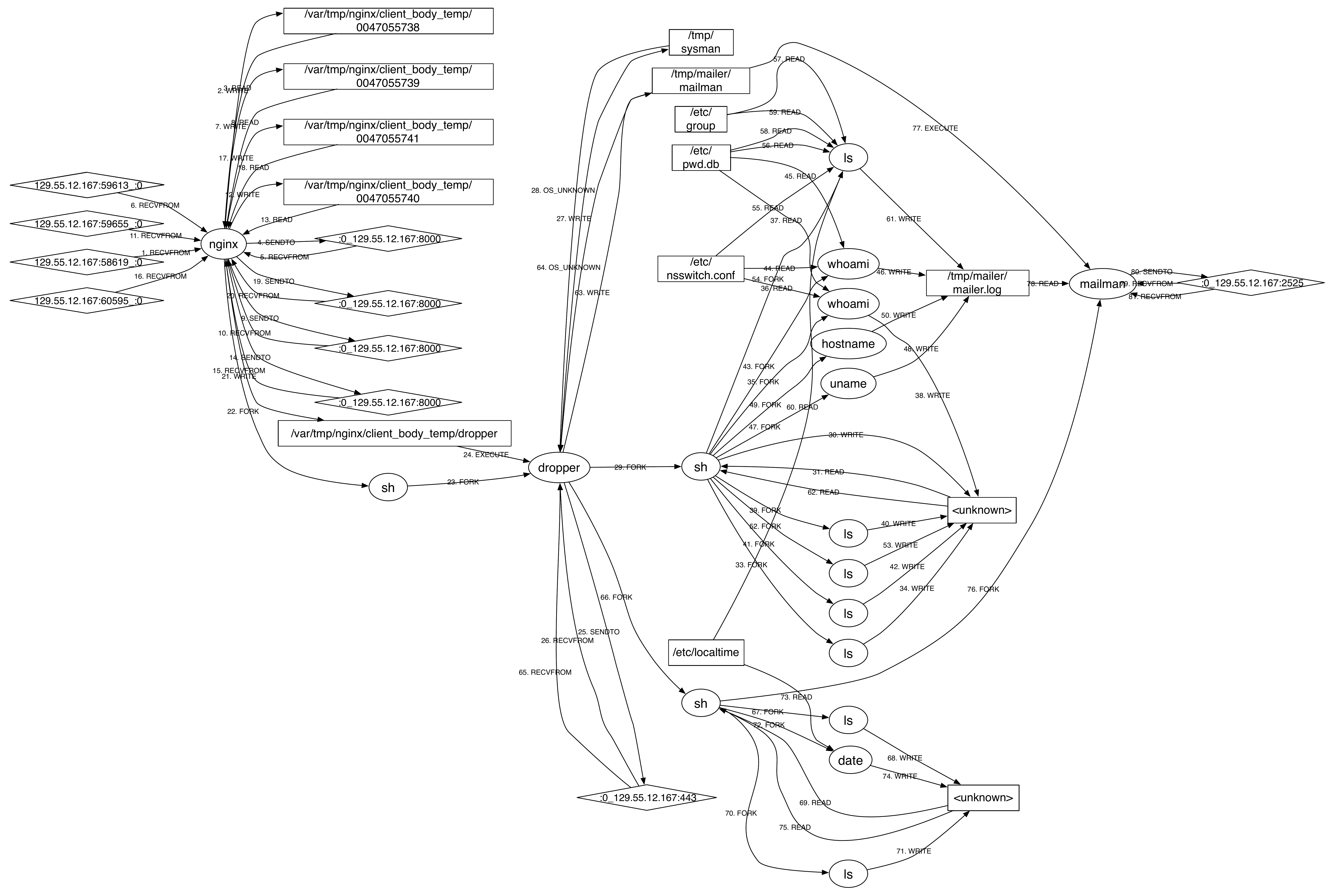}
    \caption{First Bovia Attack.}
    \label{fig:bovia1}
\end{sidewaysfigure}

In this scenario, we detected at least two instances of a similar attack, starting from the \texttt{nginx} server. In the first instance, shown in the graph in Figure \ref{fig:bovia1}, the nginx server reads from \texttt{bovia.com at 129.55.12.167:8000} (edge 20) and subsequently writes a file to \texttt{./var/tmp/nginx/client\_body\_temp/dropper}, which is then subsequently executed in a new process called \texttt{dropper} (edges 23, 24). The \texttt{dropper} process now communicates with port 443 of bovia.com and subsequently writes a file to \texttt{/tmp/sysman}. Next, the \texttt{dropper} process creates a shell (edge 29), which executes several system commands, including \texttt{ls}, \texttt{whoami}, \texttt{hostname}, \texttt{uname}. Each of these processes writes to the file \texttt{/tmp/mailer/mailer.log}. We believe that their  output  is redirected to that file. Next, \texttt{dropper} writes to the file \texttt{/tmp/mailer/mailman}, which is subsequently executed (edges 76, 77). The resulting \texttt{mailman} process reads \texttt{/tmp/mailer/mailer.log} (edge 78) and sends its contents to \texttt{129.55.12.167:2525} (edges 79, 80, 81).

\begin{sidewaysfigure}[t]
    \includegraphics[width=\textwidth]{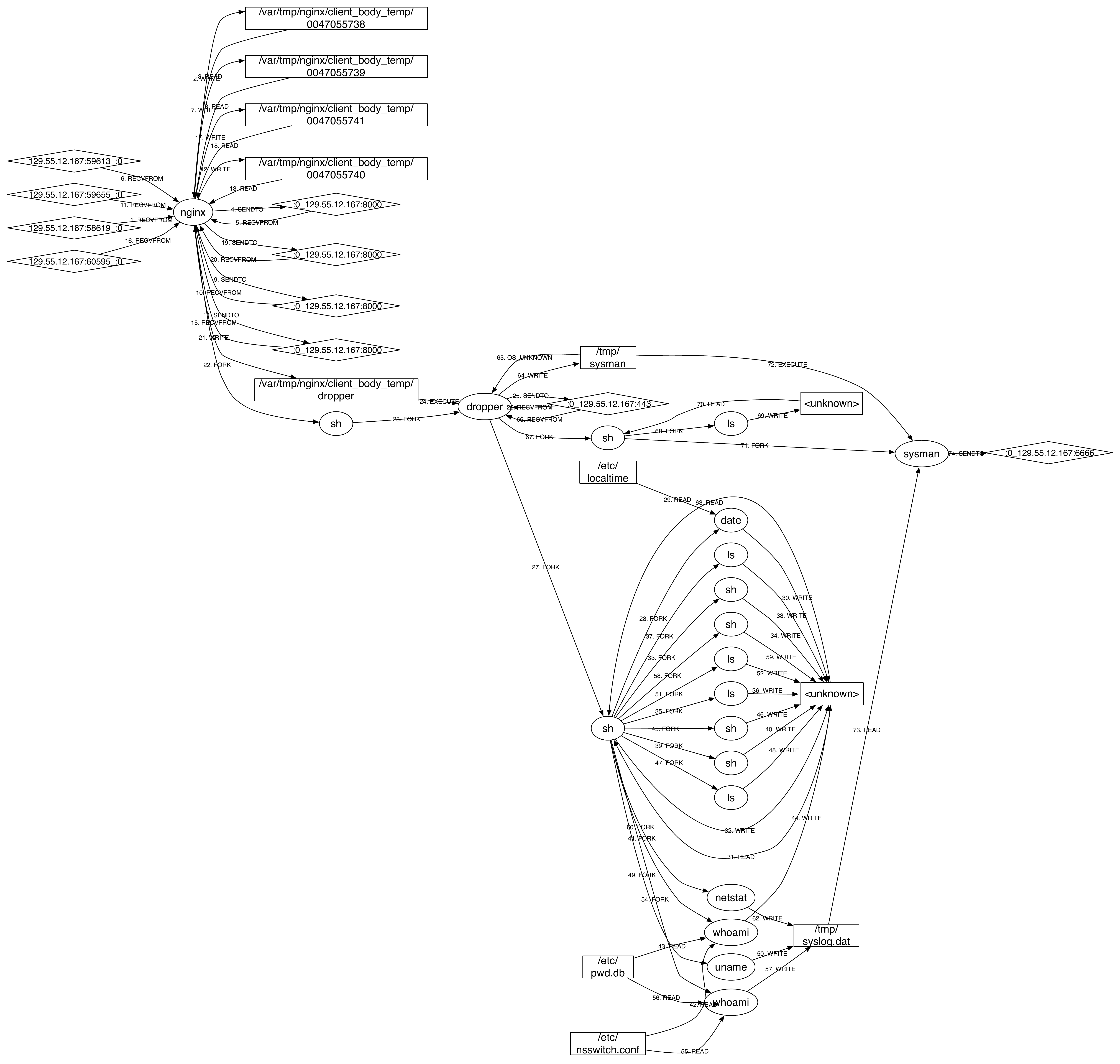}
    \caption{Second Bovia Attack.}
    \label{fig:bovia2}
\end{sidewaysfigure}

The second attack follows  a similar pattern and it is shown in Figure \ref{fig:bovia2}. The steps up to the creation of the \texttt{dropper} process are the same. However, in this second attack, dropper first writes to a file \texttt{/tmp/sysman} and then executes that file. A shell is forked from dropper as well and a series of commands are run by it, including \texttt{ls}, \texttt{whoami}, \texttt{uname}, \texttt{netstat}. The output of these commads is written to the file \texttt{/tmp/syslog.dat}, which is later read by the \texttt{sysman} process and exfiltrated to \texttt{129.55.12.167:6666}.

\subsubsection{Pandex Scenario}
\begin{sidewaysfigure}[ht]
    \includegraphics[width=\textwidth]{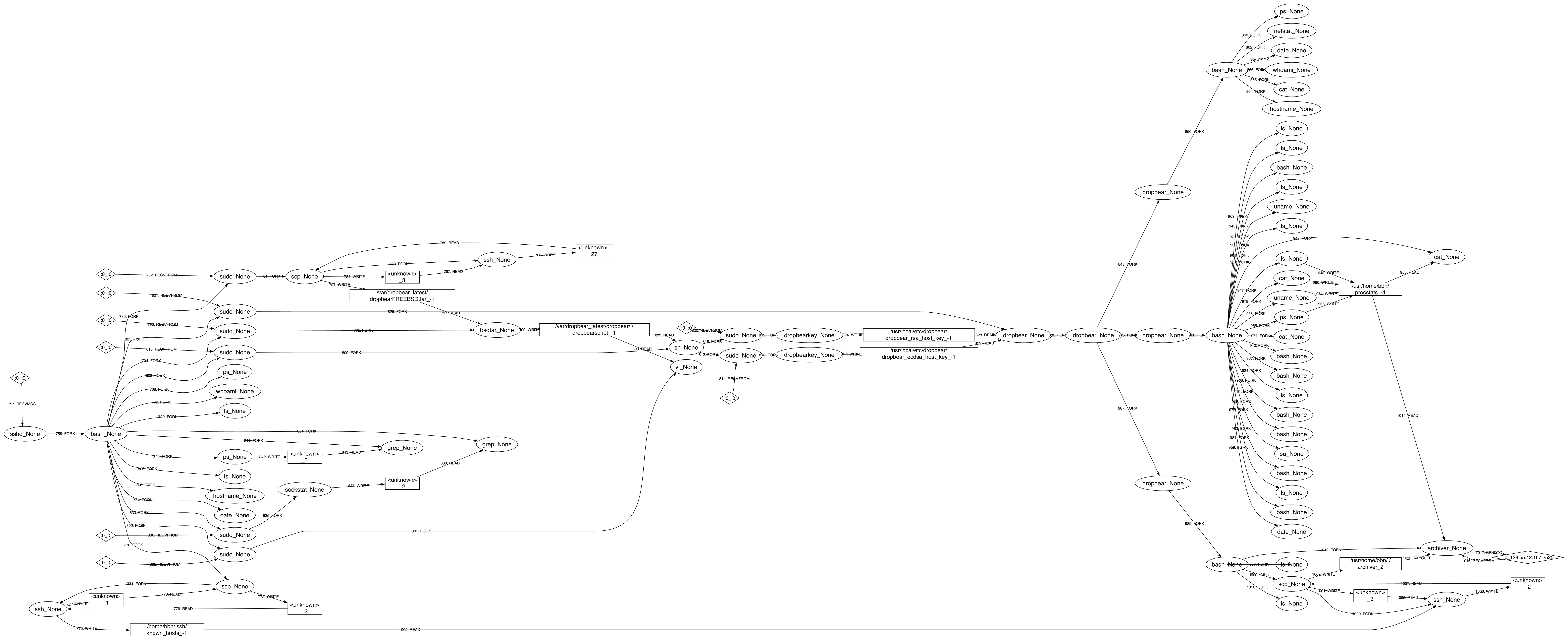}
    \caption{Pandex Attack.}
    \label{fig:pandex}
\end{sidewaysfigure}

This attack is shown in Figure \ref{fig:pandex}. The machine has an \texttt{sshd} process running. At some point \texttt{sshd}, which is possibly compromised, forks a \texttt{bash} process, we suspect under the command of the attacker (there is no evidence from the dataset of the actual attack vector, but because of the subsequent events from this shell, we conclude this is an attack). This \texttt{bash} process invokes a series of commands -- \texttt{hostname}, \texttt{whoami}, \texttt{date}, \texttt{ps}. These commands are examining the host information. 

One of the processes invoked by this bash process is \texttt{scp}, which writes to the file \texttt{/usr/home/bbn/./archiver}. This file will be executed later. We do  not see in the trace the IP address from which \texttt{scp} copies this file.

At some later point, the same \texttt{bash} process creates a (sudoed) \texttt{scp} process, which creates a new object (whose attributes are missing in the dataset) and which also downloads a file into location \\ \texttt{/var/dropbear\_latest/dropbearFREEBSD.tar}, which is then uncompressed.  

The file \texttt{dropbearscript} is next read by \texttt{sh}, and interprets it. This action creates the process \texttt{dropbearkey}, which writes  to \texttt{/usr/local/etc/dropbear/dropbear\_ecdsa\_host\_key} and \\ \texttt{/usr/local/etc/dropbear/dropbear\_rsa\_host\_key}.

Next, another \texttt{sudo} process created by \texttt{bash} starts another \texttt{dropbear} process which reads  these two keys for future use (presumably to assist in connecting to a remote host).

\texttt{Dropbear} next starts a shell, \texttt{sh}, which executes a series of commands \texttt{ls}, \texttt{bash}, \texttt{uname}, \texttt{ps}, all of which write to a file \texttt{/usr/home/bbn/procstats}. Another set of processes  are created that run \texttt{date}, \texttt{ls}, \texttt{netstat}, \texttt{whoami} and \texttt{hostname}. They write to another file (presumably a log), but the location and name attributes of this file are missing in the dataset.

In  addition, \texttt{dropbear} starts  a \texttt{bash} process, which executes the file \texttt{/usr/home/bbn/./archiver} downloaded previously via \texttt{scp}. The resulting process, called \texttt{archiver}, reads the file \texttt{/usr/home/bbn/procstats}, which contains the data output earlier, and exfiltrates this information  to \texttt{128.55.12.167:2525}.

\subsubsection{Stretch Period}
\begin{sidewaysfigure}[ht]
    \includegraphics[width=\textwidth]{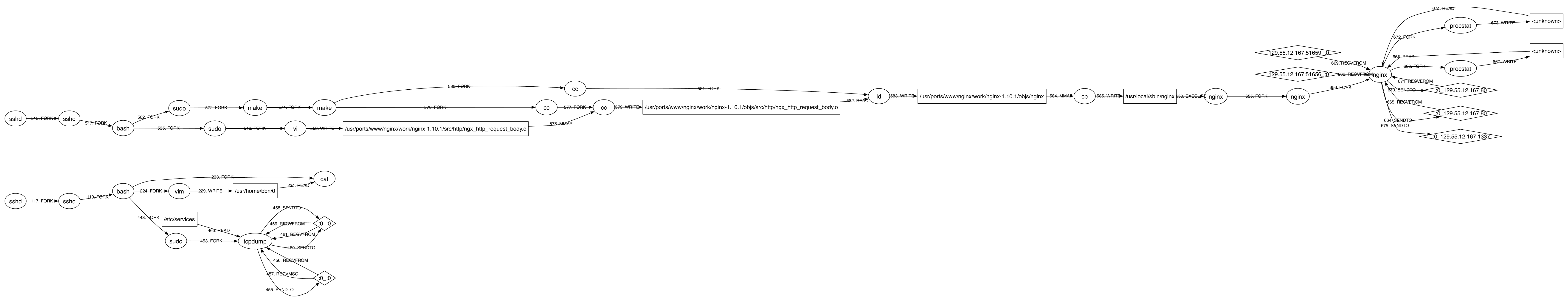}
    \caption{Stretch Period Attack (CADETS).}
    \label{fig:cadetsStretch}
\end{sidewaysfigure}

The results of the stretch period on the CADETS data re shown in Figure \ref{fig:cadetsStretch}. In particular, among several normal activities, we notice that the attack involves modifying the \texttt{nginx} server. In particular, we observed that \texttt{sshd} spawns a shell, which is the used to start \texttt{vi}. \texttt{Vi} writes to the file \texttt{ngx\_http\_request\_body.c} (edge 557). Next this file is compiled by \texttt{make}, which creates instances of \texttt{cc} and \texttt{ld}. The end result is the executable \texttt{/usr/ports/www/nginx/work/nginx-1.10.1/objs/nginx} (edges 577-583). Next this file is copied to \texttt{/usr/local/sbin/nginx} (edge 585), probably replacing the original \texttt{nginx} executable. Next,this file is executed and it reads a number of password files \texttt{/etc/spwd.db, /etc/login.conf.db}, etc. We believe that reading these files is probably legitimate behavior, since many programs routinely access those files, to deal with access rights, hence they are not shown on the graph). Next, \texttt{nginx} forks a process \texttt{procstat}, which writes to an unknown file (edges 666,667). This file is later read by \texttt{nginx} (edge 668), and such read is followed by a \texttt{SendTo} event, which sends data (presumably the file's data) to the IP address \texttt{129.55.12.167:1337}.

In addition to \texttt{nginx}, we observed a \texttt{tcpdump} process as well running on the machine. This process communicates with unknown IP addresses (probably localhost), which we could not see in the traces, and writes only to \texttt{/dev/tty}. We do not suspect this process to be malicious, unless \texttt{/dev/tty} is somehow used as a channel for data exfiltration. In the trace there is a large number of processes interacting with \texttt{/dev/tty}.

\subsubsection{Performance}
The following table documents the performance of our system on the different CADETS traces. For each data source, we report  the number of CDM records, the duration of the trace in minutes, that is the amount of tracking time, the consumption and analysis time of our system (column Analysis(mins)), as well as the main memory used by our system. As can be noted, our system can consume and analyze the data approximately 20 times faster than the speed at which they are generated (in average, under the conditions of Engagement 1).

\begin{center}
  \begin{tabular}{ | l | l |l|l| l| }
    \hline
    \textbf{Data} & \textbf{No. Records} & \textbf{Duration(mins)} & \textbf{Analysis(mins)} & \textbf{Memory(MB)} \\
    \hline
    	Bovia     & 24'448'281           & 4736						&  220                   & 872 \\
    \hline
    	Pandex    & 23'801'471			& 4744						&  198 					 & 955  \\
    \hline
    	Stretch   & 2'826'453			& 498 						&   26 					 & 125 \\
    \hline
  \end{tabular}
\end{center}

\clearpage

% =======SBU report========

%\title{\bf Analysis of FiveDirections and {\sc Trace} Data}
\section{Analysis of FiveDirections and TRACE Data: Stony Brook University}
The key elements of our approach are described below.

\subsection{Space-efficient main memory provenance graph representation}
Our subteam is focused on real-time detection and analysis capabilities. We have
found that graph databases do not provide adequate performance at the data rates
anticipated in TC. Hence we are targeting a main-memory representation of the
provenance graph. The goals of this main memory representation are:
\begin{itemize}
\item {\em High fidelity provenance tracking:} Efficiency or other considerations
  should not cause unrelated flows to be ``mixed up'' in the graph. In a
  forensic analysis, we often do not know that a particular activity was
  suspicious (or outright malicious) until a long time afterwards. If the graph
  prematurely conflates data from two sources, it would be difficult to
  disentangle the flows and accurately identify the root causes or initial
  triggers.
\item {\em Compact yet accurate representation of event data:} The
  representation should not discard event data that may be useful for anomaly
  detection, forensic analysis, or other tasks that may be performed days, weeks
  or months after events took place. Our experiments with on-disk graph
  databases has shown that graph operations would be several orders of magnitude
  slower than those on main memory data structures. However, given the volume of
  TA1 data, main memory representations would be overwhelmed unless
  ultra-compact representations are developed.
\item {\em Unified data representation across TA1s:} CDM has achieved a level
  of uniformity at the syntactic level. However, achieving semantic uniformity
  and consistency across TA1s has proven challenging because of differences in
  operating systems, instrumentation approaches, and provenance-tracking
  capabilities across TA1s. At the same time, it is not very practical or
  effective to develop (or re-implement) analysis algorithms for specific TA1s.
  For this reason, we translate TA1 data into a common semantic representation
  (CSR).
\end{itemize}
\subsection{Provenance-based attack detection}
Our approach uses provenance for real-time attack detection. Specifically, we
assign {\em integrity tags} and {\em confidentiality tags} to objects (files, IPCs,
etc.) and subjects (processes). In the simplest case, there are two levels of
integrity: {\em benign,} representing data and code from sources that
are trusted to be non-malicious, and {\em untrusted,} representing code/data
from anywhere else. Similarly, two level of confidentiality can be defined in
the simplest case: {\em public} and {\em private.}

Pre-existing objects and subjects are assigned initial tags. Newly created
subjects inherit the tags of their parents.  Newly created objects
inherit the tags of the process creating them. In addition, tags flow with
data. For instance, a benign subject will acquire {\em untrusted} tag if it
reads an object with untrusted tag. Based on these tags, we can raise alarms on
several suspicious behaviors:
\begin{itemize}
\item {\em Untrusted code execution:} This alarm is raised when a subject with
  higher code integrity loads or executes untrusted code.
\item {\em Undesired Subject/Object Downgrading:} This alarm is raised when
  a higher integrity subject or object is downgraded because of a flow from a
  source of lower integrity. This class captures attacks where (a) a benign
  process is exploited by untrusted input, or (b) a compromised process corrupts
  a benign object\footnote{Note that objects here need not be files, but can be
    intranet sites as well. So, it should capture the example of a malicious web
    site that contains JavaScript that accesses an intranet server. This assumes
    that the top-level browser does not read any untrusted site, and only
    subprocesses/units/tabs do, and that these tabs run the malicious JS in
    question.}.
\item {\em Confidential data leak:} An alarm can be raised when untrusted
  subjects acquire (i.e., read) or exfiltrate (i.e., write) sensitive data.
\item {\em Permission changes:} When suspicious changes to file permissions
  occur, e.g., when an untrusted object is made executable, an alarm can be
  raised\footnote{Note that this policy is applicable to file objects, as well
    as in-memory objects. Normal {\tt mmap}/{\tt mprotect}s that occur in
    conjunction with file loading are supposed to be recognized by the input
    stage, which instead reports a loadlib event.}.
\end{itemize}
In our implementation, several additional levels of integrity and
confidentiality are used. Moreover, tag assignment, propagation and
attack detection are all controlled by {\em provenance policies} that are
can tuned by a forensic analyst.
\subsection{Tag-based forward analysis} \label{fwdanal}
This analysis is initiated when an alarm is raised by one of the policies
described in the previous section. All subsequent actions performed by the
subject involved in the alarm, as well as its descendants, are emitted by the
system, and can be used to construct a graph of subsequent activities. In order
to fine-tune the subgraph that is emitted, we rely on the customizable
policy framework mentioned in the last section.  This allows us to ``tune out''
uninteresting events, such as loads or reads of  files with high integrity and
low confidentiality.

In the first engagement, most of our successful analyses were based on the {\em
  untrusted code execution} alarm. A secondary policy was defined to cope with
instances where provenance is not accurately captured in TA1 data. In such a
case, an object may be written, but we may not have the correct provenance, and
hence its tag may be incorrect. Our ``backup'' policy to deal with such
instances was to flag execution of a previously written file.

\subsection{Performance}
\subsubsection{Size}
Figure~\ref{memuse} shows the sizes and memory used for the data sets we
consumed and successfully analyzed. In summary, we make the following
observations:
\begin{itemize}
\item We first translate CDM into our common semantic representation (CSR),
  a uniform representation that we use across different OSes. CSR is stored
  in compressed (gzipped) format. On average, CSR is about
  40 times smaller than the binary representation used in CDM.
\item On average, we use about one byte of memory per CDM record for
  data from FiveDirections, and about 4 bytes per CDM record for data from {\sc Trace}.
\end{itemize}

\begin{figure}[h]
  \begin{center}
  \begin{tabular}{||c|r|r|r|r|r||}
    \hline
    Data set & CDM  & CSR  & CSR/CDM & Main & Memory/ \\
             & size &  Size & ratio & memory& CDM ratio \\
    \hline \hline
    5D Bovia & 150 MB  & 3 MB&  2.0\% & 2.1 MB & 1.4\%\\ \hline
    5D Pandex & 130 MB & 2.5 MB& 1.9\% & 1.7 MB & 1.3\%\\ \hline
    5D Stretch &  49 MB & 0.8 MB& 1.6\% & 0.6 MB & 1.2\%\\ \hline \hline
    \multicolumn{3}{||c}{\bf 5D Mean}  & {\bf 1.9\%} & \multicolumn{2}{|r||}{\bf 1.3\%}\\ \hline\hline
    {\sc Trace} Bovia & 175 GB & 4.7 GB& 2.7\% & 11.3 GB & 6.4\%\\ \hline
    {\sc Trace} Pandex & 111 GB & 3.9 GB& 3.5\% & 5.8 GB & 5.2\%\\ \hline
    {\sc Trace} Stretch & 13 GB & 0.4GB& 3.1\%& 0.7 GB & 5.4\%\\ \hline \hline
    \multicolumn{3}{||c}{{\bf {\sc Trace} Mean}}    & {\bf 3.1\%} &  \multicolumn{2}{r||}{\bf 5.7\%}\\ \hline
  \end{tabular}
  \end{center}
  \caption{Size of main memory provenance graph and common semantic
    representation (CSR). ``CDM size'' refers to the size of CDM binary format,
    as stored in Kafka queues. CSR is stored in gzipped text format.} \label{memuse}
\end{figure}
%\pagebreak
\subsubsection{Runtime}
\begin{figure}[h]
  \begin{center}
  \begin{tabular}{||c|r|r|r|r|r|r||}
    \hline
    Data set & CDM  & Duration & \multicolumn{2}{|c|}{Graph Construction}  & \multicolumn{2}{|c||}{Analysis} \\
    \cline{4-7}
             & &  & Time & Speed-up & Time & Speed-up\\
    \hline \hline
    5D Bovia & 150 MB  & 19:43:46 & 0.92s & 77 K & 1.05s & 68 K\\ \hline
    5D Pandex & 130 MB & 23:15:42 & 0.75s&  112 K & 0.86s & 97 K\\ \hline
    5D Stretch &  49 MB & 06:22:42 & 0.24s& 96 K & 0.27s & 85 K\\ \hline \hline
    \multicolumn{4}{||c}{\bf 5D Mean} & {\bf 95 K} & \multicolumn{2}{|r||}{\bf 83 K}\\ \hline\hline
    {\sc Trace} Bovia & 175 GB & 79:06:39 & 1.22h & 65 K & 2.5h & 32 K\\ \hline
    {\sc Trace} Pandex & 111 GB & 79:05:13 & 0.77h & 102 K & 1.73h & 46 K\\ \hline
    {\sc Trace} Stretch & 13 GB & 07:59:26 & 333s & 86 K & 375s & 77 K\\ \hline \hline
    \multicolumn{4}{||c}{\bf {\sc Trace} Mean}  & {\bf 84 K} &  \multicolumn{2}{|r||}{\bf 52 K}\\ \hline
  \end{tabular}
  \end{center}
  \caption{Runtime performance for main memory provenance graph construction and
    analysis. ``CDM size'' refers to the size of CDM binary format, as stored in
    Kafka queues. The data collection period is shown as ``duration,'' in the
    format HH:MM:SS. Graph construction and analysis times are shown in either
    seconds or hours. ``Speed-up'' is given by duration/time. It represents
    the average number of simultaneous data streams that can be handled by our
    system while using {\em a single core of a single processor.} Performance
    measurements were made on a 2.8GHz AMD Opteron 62xx processor,
    with 48GB main memory, and running Ubuntu 16.04.} \label{cpuuse}
\end{figure}
\subsubsection{Analysis Selectivity and Automation}
In this section, we focus on how much automation is brought to the forensic
analysis by our techniques. In particular, attack detection is automated, and
it identifies the most suspicious events. Starting from these events, our system
performs an automated forward analysis. During this analysis, configurable
policies are used to filter out uninteresting events performed by suspect
processes (e.g., loads of benign libraries), while retaining significant events.
This results in a very small fraction of the CDM events being flagged for
further analysis.

In some cases, the filtered event set is small enough to be directly fed into a
graph visualization program such as {\tt graphviz}. In other cases, the number
of events is too large, and is subject to further manual analysis.
Figure~\ref{selectivity} shows how selective these automated processes are. Less
than one-thousandth of the CDM events are flagged for further analysis.

\begin{figure}[h]
  \begin{center}
  \begin{tabular}{||c|r|r|r||}
    \hline
    Data set & Number of & Number of  & Number of \\
             & CDM records & Filtered Events & Alarms\\
    \hline \hline
    5D Bovia & 1.85 M & 1.6 K & 3 \\ \hline
    %5D Pandex & 509 K & 23:15:42 & 0.75s\\ \hline
    5D Stretch &  509 K & 0.18 K & 4 \\ \hline \hline
    {\sc Trace} Bovia & 2.3 B & 28 K & 12 \\ \hline
    {\sc Trace} Pandex & 1.5 B & 25 K & 6 \\ \hline
    {\sc Trace} Stretch & 168 M & 4 K & 4 \\ \hline \hline
  \end{tabular}
  \end{center}
  \caption{Filtered events and alarms generated for further analysis} \label{selectivity}
\end{figure}

Even in those cases where the approach is selective enough to be directly
fed into a graph visualization tool, the result of such an
automated process will still include some details that are not critical for
understanding a scenario. Since one of our goals in this report is to provide
graph representations that are easy for TA 5.1 to understand, we performed further interactive filtering and editing (of the above filtered events) to
arrive at the graphs shown in the next two sections. 
\subsection{Analysis of FiveDirections Data}
\subsubsection{Issues and Challenges}
We successfully consumed data from FiveDirections. All three data sets
(Bovia, Pandex and Stretch) were successfully consumed. Traces of Bovia and
Pandex are divided into multiple topics because of crashes of their system. We
are able to reconstruct the Bovia scenario and Stretch scenario. We analyzed
the Pandex scenario but there is nothing suspicious in the provided traces.

We encountered several significant difficulties with this data. First, network
flows are encoded in CDM in an unusual way that made it difficult to process
them correctly, at least in the beginning. This is an instance of the more
general problem that CDM provides a syntax without nailing down the semantics.
Second, there is an inconsistency in how network I/O is handled, as compared to
file I/O. In the case of network connections, only the initial connection
operation is reported, but none of the subsequent reads or writes. In contrast,
reads and write operations are reported on files, but the open operation goes
unreported.

Other difficulties specific to FiveDirections are of greater concern to us
going forward. First, access to devices such as the keyboard, camera and
microphone are not captured by FiveDirections. Second, activities of system processes
 seem not to be captured in some cases. The Pandex scenario involves RDP, and
it appears that activities initiated through RDP are not recorded by Five
Directions. This obviously makes it extremely difficult, if not impossible, to
do any meaningful analysis of their Pandex data.
\begin{figure}
\begin{center}
\vspace*{-2em}
\hspace*{-.7in}\includegraphics[width=1.17\textwidth]{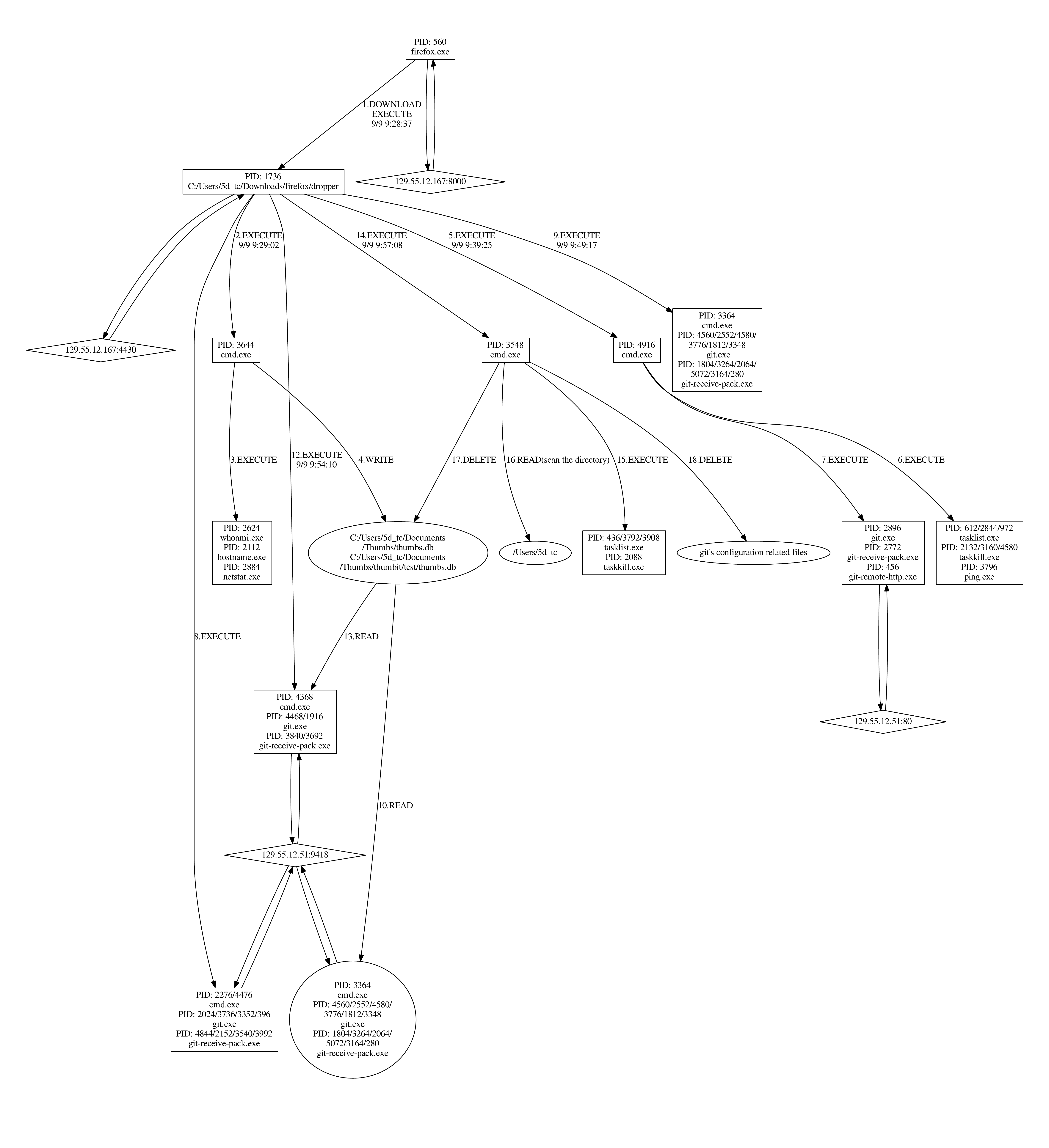}
\vspace*{-6em}
\end{center}
\caption{FiveDirections: Bovia Scenario (1 of 3)}
\label{fdbov1}
\end{figure}

\subsubsection{Bovia Scenario}
This analysis applies to what we found in the fifth (last) Kafka queue for FiveDirections
Bovia data. The starting point of our analysis are the following two alarms
generated by our system:\medskip

\begin{footnotesize}
  \begin{tt}
\noindent
16-09-09 09:28:37.832: Alarm: UntrustedLoad: Object \verb+C:\Users\5d_tc\Downloads\firefox\dropper+\linebreak\indent
Subject pid=560 \verb+C:\Program Files\Mozilla\firefox\firefox.exe+\\[0.7ex]
16-09-09 09:55:07.049: Alarm: UntrustedLoad: Object \verb+C:\dropper+\linebreak\indent
Subject pid=1580 \verb+C:\Program Files\Mozilla\firefox\firefox.exe+
\end{tt}
\end{footnotesize}

\begin{figure}
\begin{minipage}{2.5in}
\begin{center}
%\vspace*{-2em}
\includegraphics[width=2.5in]{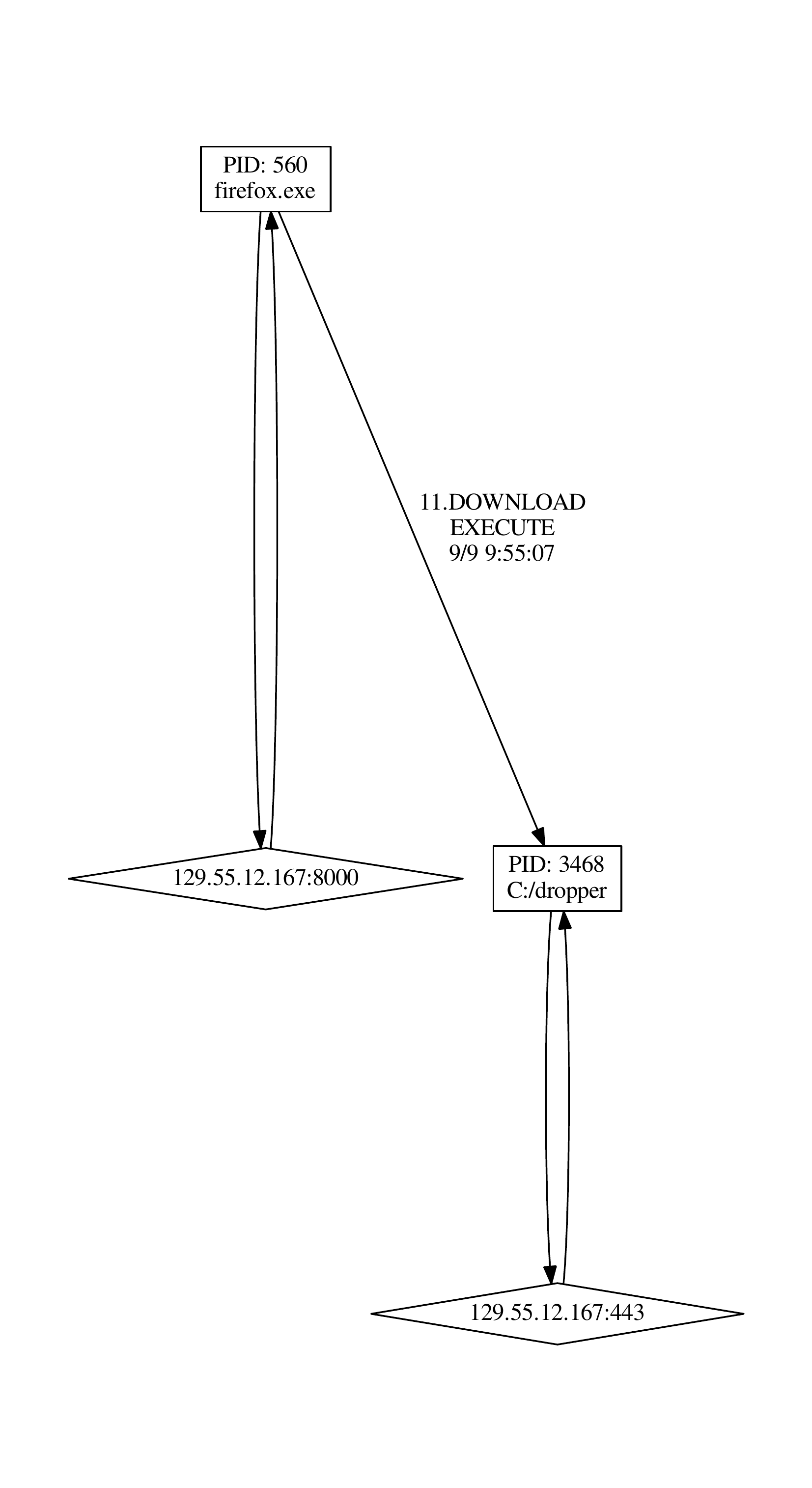}
%\vspace*{-4em}
\end{center}
\end{minipage} \hfill
\begin{minipage}{3.5in}
\begin{center}
%\vspace*{-2em}
\includegraphics[width=3.8in]{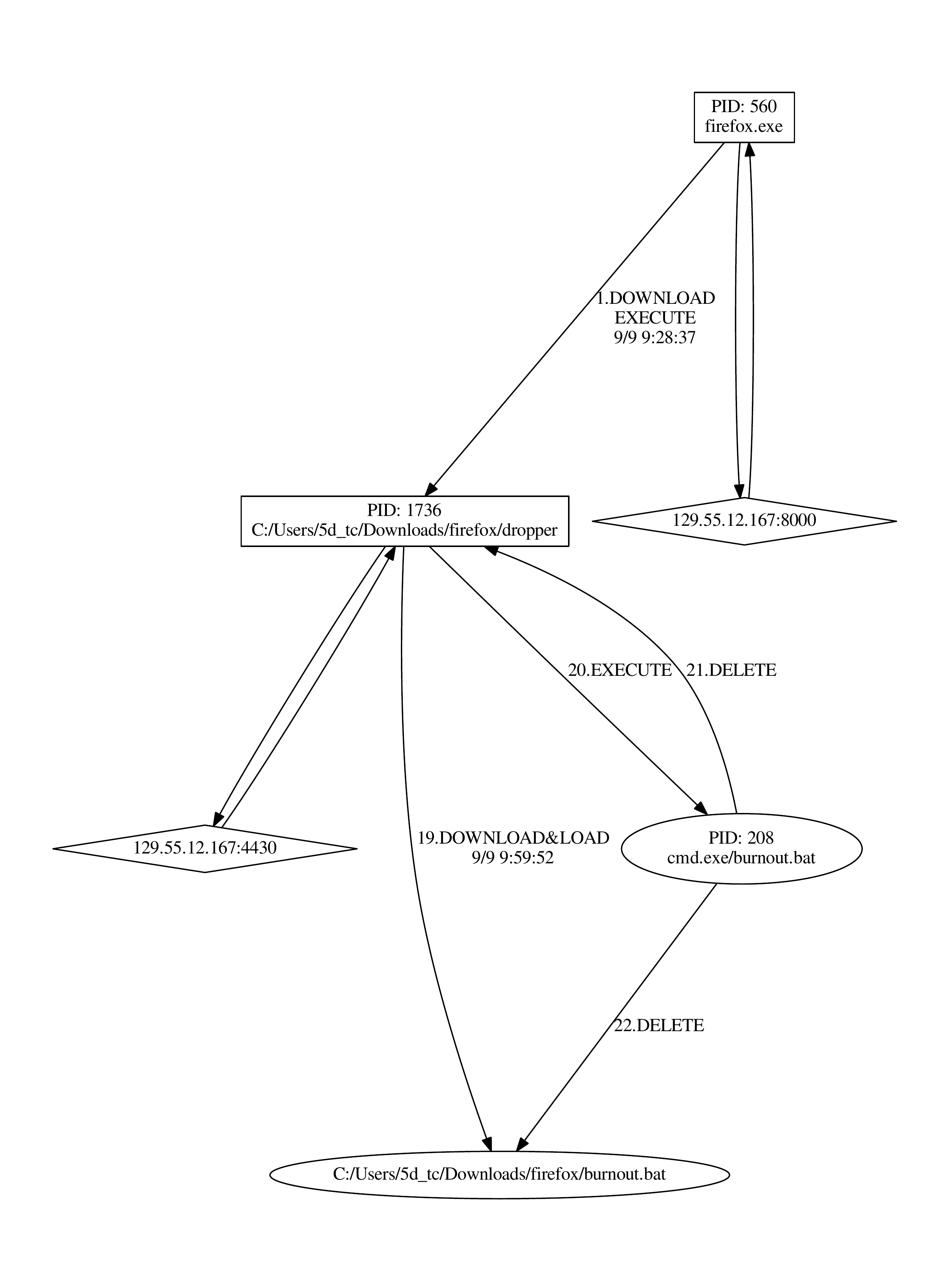}
%\vspace*{-4em}
\end{center}
\end{minipage}
\caption{FiveDirections: Bovia Scenario (2 and 3 of 3)} \label{fdbov3}
\end{figure}

\noindent
Using forward analysis (as sketched in Section~\ref{fwdanal}) from the point of
the above alarms, we arrived at the following reconstruction of the attack. The
attack lasts for about half an hour. During this period, the attacker seems to
engage in 12 interactive sessions over which the attacker invokes over 60 commands
that end up executing about 6K to 7K system calls. Figures~\ref{fdbov1}--\ref{fdbov3} illustrates the steps.

The entry point for the attack is Firefox, which seems to be compromised when
visiting the web server at 129.55.12.167 on 9/9/16 at 9:28am. There are connections
to this web server from about 3 minutes before this point, with the difference
that the last connection before the compromise goes to port 8000 rather than 80.
(Because of the lack of granularity in FiveDirections data, we cannot attribute the network
source exactly; instead we are hypothesizing that the last connection before the
compromise is mostly likely responsible.)

When Firefox is compromised, a malicious program called {\tt dropper} is
downloaded. Firefox then executes dropper. Dropper then invokes {\tt cmd.exe} 12
times over the next half hour, using it to perform various data gathering and
exfiltration tasks, as described further below. Dropper is likely providing a
remote interactive shell capability, connecting to 129.55.12.167:4430 to receive
commands and executing them via {\tt cmd.exe} each time.

Some of the {\tt cmd.exe} sessions last less than a minute, while others last
for maybe 10 minutes or more. Some of these sessions invoke {\tt
  PowerShell.exe}. Since there are so many {\tt cmd.exe} sessions, we will
summarize the overall activity, instead of discussing each session in detail.

These various {\tt cmd.exe} sessions collect data in a file called {\tt thumbs.db},
which is then exfiltrated using git to 129.55.12.51:9418. There is some activity
by {\tt cmd.exe} that looks like an attempt to reconfigure {\tt git} to connect
to the malicious server in the future. Towards the end of the attack, several of
{\tt git}'s configuration related files are cleaned up, presumably to
clean up the system and avoid leaving any trails.

Given that the attacker is able to freely access the files used by {\tt git},
and what appears to be the project directory ({\tt Thumb}), the attacker could have
modified these documents or source code, and pushed them to the repository.
However, we cannot find any evidence that this was actually done, as we don't
see source code files or meaningful document files being edited or modified in
another way.

In one of the sessions, information gathering commands such as {\tt whoami},
{\tt hostname} and {\tt netstat} are run, and the results are saved into a file called
\verb+C:\Users\5d_tc\Documents\Thumbs\thumbs.db+. Sometimes the name changes to
\verb+C:\Users\5d_tc\Documents\Thumbs\thumbit\test\thumbs.db+. In the end, both
files are removed in order to hide the attacker's trail.

Another session seems to scan all of the directories under \verb+\Users\5d_tc+.
This could be a case of an attacker interactively examining the system. We
cannot find any sign of the resulting data being saved to a file or
exfiltrated.

At the very end, dropper downloads a batch file called {\tt burnout.bat} into
Firefox download directory and then executes it. The purpose of this batch file
seems to be to delete dropper as well as {\tt burnout.bat} from the system, thus
removing all signs of a break-in.

\subsubsection{Pandex Scenario}
We were told that RDP played a central role in the Pandex attack. Based on our
discussions with FiveDirections, it appears that activities initiated through
RDP are not recorded by Five Directions. This obviously makes it extremely
difficult, if not impossible, to do any meaningful analysis of their Pandex
data.

We fell back on our ``backup'' detection technique, looking for files that
are written during the engagement and then subsequently loaded or executed.
Based on our understanding of the rules of engagement, these attacks involve, at
some point, downloading a file that is then loaded or executed. While the
mechanism for downloading may differ (e.g., download via browser, email, or a
remote login session; download of an executable, DLL, or a file taken as input
by an interpreter), we did expect there to be malicious executable content that
is saved on the disk.

However, we found that none of the files written during the
engagement were loaded or executed, not even by interpreters. In fact,
we looked at the set of all files written on any of the four Kafka queues
for FiveDirections Pandex, and compared it with the list of files executed/loaded
across them, and again found no intersection. This could, of course, be the
result of above-mentioned gaps in FiveDirections data --- files written
as a result of an RDP session may not be captured in the data.

As a last resort, we looked at the list of all programs and DLLs loaded or
executed in the last Kafka queue for Pandex scenario, and did not find even
a single suspicious-looking process.

While our reasons for missing data are speculative, it seems safe to say that
there is minimal, if any, trace of attack in this data.
\subsubsection{Stretch Scenario}
The starting point for our analysis are the following four alarms from one of
our detectors. (We enabled only one of our detectors that detects untrusted code execution.)\medskip

\begin{figure}
\begin{center}
\vspace*{-2em}
\includegraphics[width=4in]{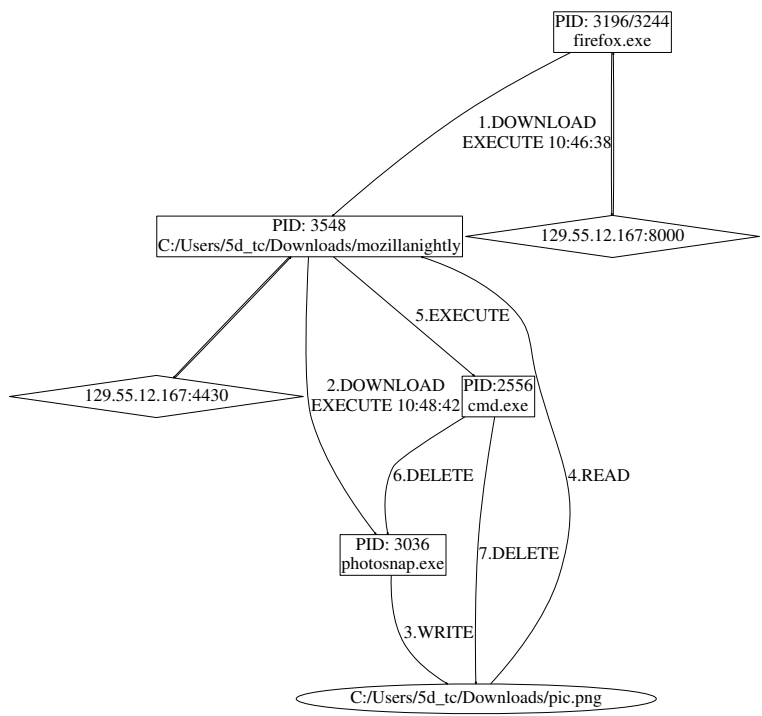}
\vspace*{-2em}
\end{center}
\caption{FiveDirections: Stretch Scenario (1 of 2)}
\label{fdstr1}
\end{figure}

\begin{figure}
\begin{center}
\vspace*{-3em}
\includegraphics[width=\textwidth]{{sbu-figs/5dstretch2}}
\vspace*{-3.7em}
\end{center}
\caption{FiveDirections: Stretch Scenario (2 of 2)}
\label{fdstr2}
\end{figure}

\begin{footnotesize}
  \begin{tt}
\noindent
16-09-27 10:46:38.093: Alarm: UntrustedLoad: Object \verb+C:\Users\5d_tc\Downloads\mozillanightly+\linebreak\indent Subject pid=3244 \verb+C:\Users\5d_tc\Downloads\firefox\firefox.exe+\\[.8ex]
16-09-27 10:48:42.855: Alarm: UntrustedLoad: Object \verb+C:\Users\5d_tc\Downloads\photosnap.exe+\linebreak\indent Subject pid=3548 \verb+C:\Users\5d_tc\Downloads\mozillanightly+\\[.8ex]
16-09-27 10:48:54.458: Alarm: UntrustedLoad: Object \verb+C:\Users\5d_tc\Downloads\mozillanightly+\linebreak\indent Subject pid=3404 \verb+C:\Users\5d_tc\Downloads\firefox\firefox.exe+\\[.8ex]
16-09-27 14:22:12.615: Alarm: UntrustedLoad: Object \verb+C:\Users\5d_tc\Downloads\firefox\mozillanightly+\linebreak\indent Subject pid=3196 \verb+C:\Users\5d_tc\Downloads\firefox\firefox.exe+
\end{tt}
\end{footnotesize}

\noindent
From the alarms, it is clear that  two versions of {\tt mozillanightly}
malware are being downloaded. The first version is executed twice, while the second
version is executed once. Using forward analysis (as sketched in
Section~\ref{fwdanal}) from the point of the above alarms, we arrived at the
following reconstruction of the attack. Figures~\ref{fdstr1} and \ref{fdstr2}
illustrate the steps of this attack.

The attack begins with what seems to be a browser compromise of Firefox at
10:45:38, which causes a malicious executable {\tt mozillanightly} to be
downloaded from IP address 129.55.12.167:8000. There is a second download of
malware with the same name at 14:22:12. These are stored in two
different places, as shown in Figures~\ref{fdstr1} and \ref{fdstr2}.

At 10:47, Firefox executed \verb+C:\Users\5d_tc\Downloads\mozillanightly+, which
downloads \verb+photosnap.exe+ from 129.55.12.167:4430. Then {\tt
  mozillanightly} executes {\tt photosnap.exe}. Since FiveDirections does not
record device access, there is no direct information about the input to {\tt
  photosnap.exe}. However, from the DLLs it loads, we surmise it is either
taking a picture using the camera or performing a screengrab. The results are
saved in a couple of temporary files ({\tt img57.tmp} and {\tt img58.tmp}).
Subsequently, the contents of these files seem to be copied over into
\verb+C:\Users\5d_tc\Downloads\pic.png+, and then this file is sent out to
129.55.12.167:4430. At last, {\tt mozillanightly} invokes {\tt cmd.exe} to remove {\tt
  pic.png} and {\tt photosnap.exe}.

At 10:49, Firefox executes \verb+C:\Users\5d_tc\Downloads\mozillanightly+ again,
but nothing special happens in this session. Perhaps the attacker was
interactively examining the system, running commands such as {\tt hostname} and
{\tt whoami}, but no files seem to have been created or exfiltrated.

At 14:32, Firefox executes
\verb+C:\Users\5d_tc\Downloads\firefox\mozillanightly+ again, which in turn
invokes {\tt cmd.exe} to run {\tt hostname.exe}, {\tt whoami.exe} and {\tt
  netstat.exe}. The results of these commands are written to a file named
\verb+C:\Users\5d_tc\Downloads\firefox\4662.log+.

Then {\tt mozillanightly} downloaded
\verb+C:\Users\5d_tc\Downloads\firefox\mnsend.exe+ from 129.55.12.167:4430, and it
invoked {\tt cmd.exe} to execute {\tt mnsend.exe} to send {\tt 4662.log} to
129.55.12.167: 7770, and then it removed {\tt 4662.log} and {\tt mnsend.exe}.

At last, {\tt mozillanightly} downloads
\verb+C:\Users\5d_tc\Downloads\firefox\burnout.bat+ from 129.55.12.167:4430. When
    {\tt burnout.out} executes, it completes the clean up after the attack,
    removing \verb+C:\Users\5d_tc\Downloads\firefox\mozillanightly+ and {\tt
      burnout.bat}.
\subsection{Analysis of {\sc Trace} Data}
\subsubsection{Issues and Challenges}
We have successfully consumed all three data sets from (Bovia, Pandex and
Stretch). We were able to reconstruct all three scenarios and detect the
attacks. Consuming the data turned out to be challenging as there were some
changes of semantics in the data representation during the engagement compared with the data
that we got previously. For example, CREATE events edges that contained file uuid
information were switched between toUuid and fromUuid during the engagement.

There were 60 times more memory objects than all the file and netflow objects
combined, which created an explosion in the number of events related to them.
For example there were 276 million memory objects in the Bovia scenario, even a
single event occurrence on each memory object leads to generate approximately 1 billion
records (definition of the memory object, event definition and 2 edges), which
is almost half the records in the scenario.

Instead of handling stdin, stdout and stderr, they were reported as SrcSink
objects with the only indication that they were indeed stdin, stdout and stderr
being the file descriptor. Also, there were pipes which were read from but were
never written to, which leads us to think of missing information flow from the
program's stdout to pipes. Also, the case where pipes were written to and never
read from was seen. This was seen in the attack scenario, in the stretch data,
where the program that performed cat operation on the passwd file writes to a
pipe and an openssl program also writes to that same pipe, but that pipe was never read
from. As a result, this data does not tell us where the information went.
Other issues include:
\begin{itemize}
\item subjects that did not have proper clone and execve sequences
\item multiple unlink operations on the same object
\item netflow objects with no IP addresses, and
\item declaration of unused subjects/objects.
\end{itemize}
The only way we found to consume the data was to drop events affected by these
issues and also drop SrcSink objects and related events except for the
ones related to stdin, stdout and stderr. As a result, significant provenance
information may have been lost.
\subsubsection{Bovia Scenario}
\begin{figure}
\begin{center}
\vspace*{-2em}
\includegraphics[width=0.92\textwidth]{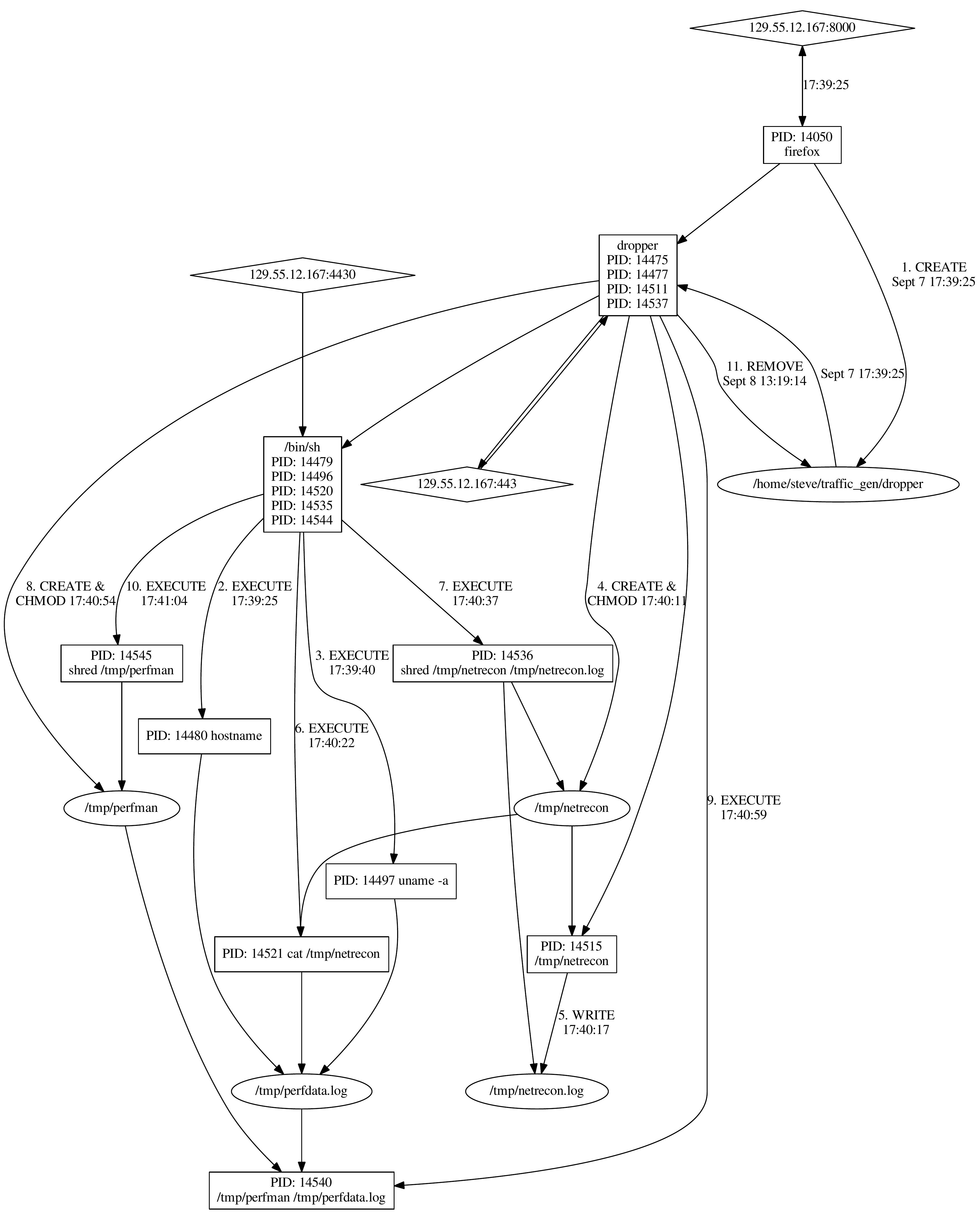}
\vspace*{-2em}
\end{center}
\caption{{\bf {\sc Trace} Bovia Scenario: Episode 3.} This episode also begins
  with the download of {\tt /home/steve/traffic\_gen/dropper} from
  129.55.12.167:8000 and its execution on 9/7 at 17:39:25. Next, the attacker
  executes the programs {\tt hostname} and {\tt uname} {\tt -a}, with output
  written to {\tt /tmp/perfdata.log}. The dropper process downloads {\tt
    /tmp/netrecon} and performs {\tt chmod} on it. Next, {\tt netrecon} is
  executed, which results in the creation of {\tt /tmp/netrecon.log} file. The
  attacker performs a {\tt cat} on {\tt netrecon} file and the output is written
  to {\tt perfdata.log}. Next, {\tt shred} is run on {\tt netrecon} and {\tt
    netrecon.log}, presumably to destroy evidence of the break-in. Then the
  dropper process downloads {\tt /tmp/perfman} file and performs {\tt chmod} on
  it. Next, {\tt perfman} is executed with {\tt perfdata.log} file, and then
  {\tt shred} is run on {\tt perfman}. On the next day at 13:19:14, the dropper
  file is deleted.
}
\label{sribov3}
\end{figure}

\begin{figure}
\begin{center}
\vspace*{-2em}
\includegraphics[width=0.9\textwidth]{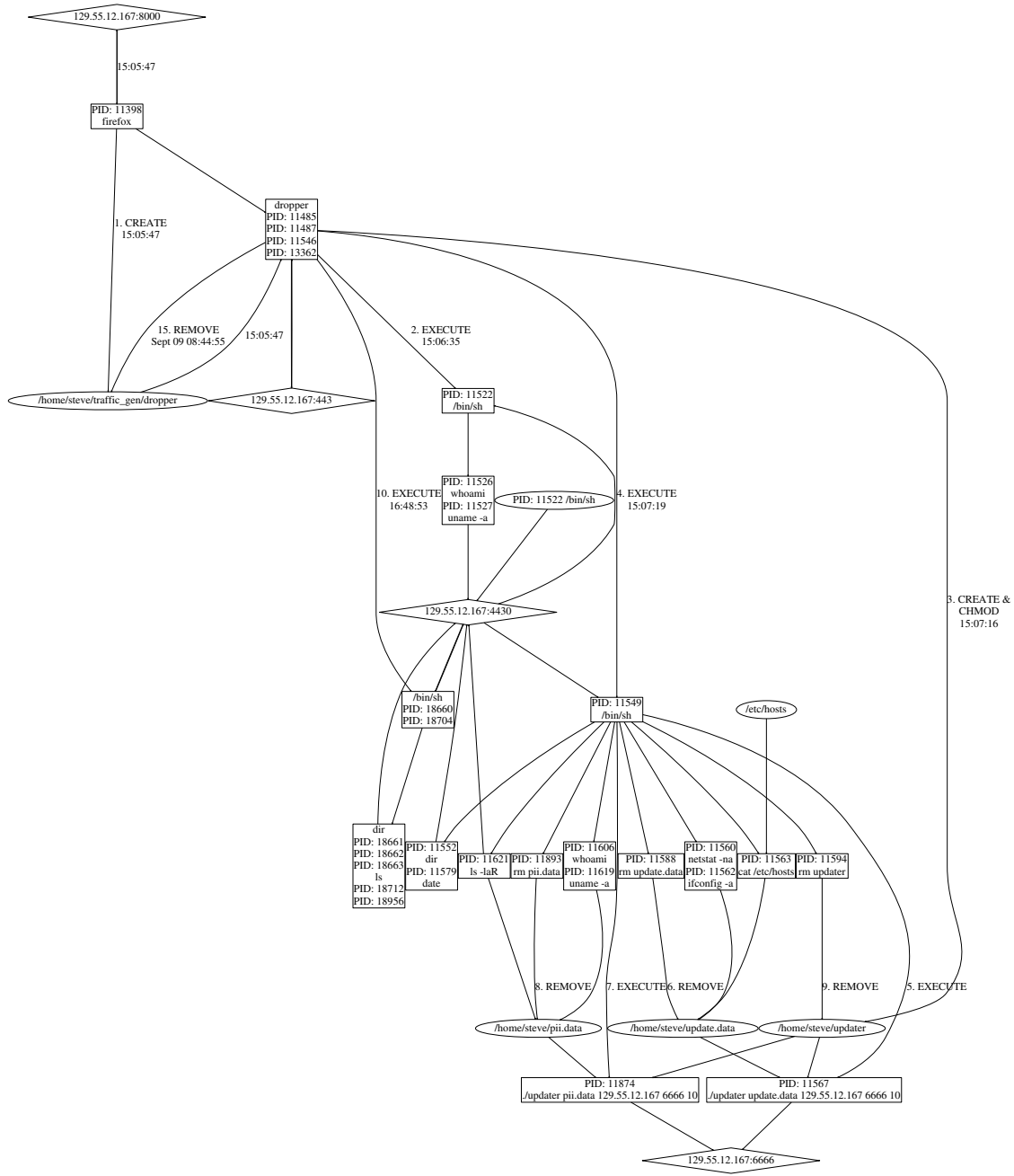}
\vspace*{-2em}
\end{center}
\caption{{\bf {\sc Trace} Bovia Scenario: Episode 4 (Part 1).}
  This episode begins with the download of {\tt /home/steve/traffic\_gen/dropper} from 129.55.12.167:8000 and its execution on 9/8 at 15:05:47. The attacker gathers information via {\tt uname} {\tt -a} and {\tt whoami} and sends it through 129.55.12.167:4430. The {\tt dropper} program creates another file {\tt /home/steve/updater} and performs {\tt chmod} on it. More information is gathered using
  {\tt netstat} {\tt -na} and {\tt ifconfig} {\tt -a}, and is written to the file {\tt /home/steve/update.data}. The updater program is then run with {\tt update.data} file and the information is sent to 129.55.12.167:6666. After this,
  {\tt update.data} is deleted.   The attacker runs the programs {\tt uname} {\tt -a}, {\tt whoami},
  {\tt ls} {\tt -laR} and the output is written to the file {\tt /home/steve/pii.data}. The updater program is then run with {\tt pii.data} and the information is sent to 129.55.12.167:6666. After that the file {\tt pii.data} is deleted. Next the updater program is deleted. The attacker next checks the directory by running {\tt ls} and {\tt dir}.
}
\label{sribov41}
\end{figure}

\begin{figure}
\begin{center}
\vspace*{-4em}
\includegraphics[width=\textwidth]{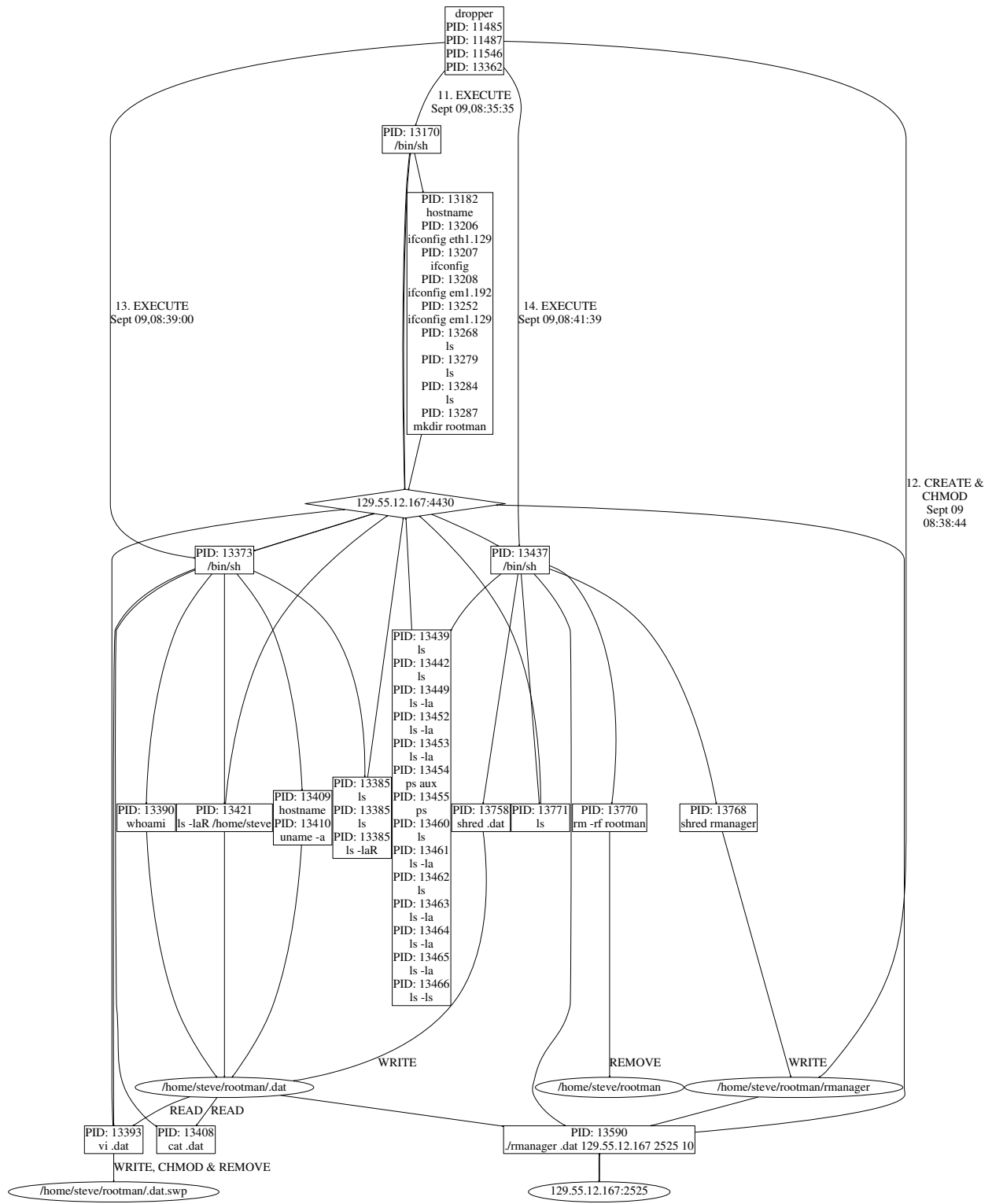}
\vspace*{-2em}
\end{center}
\caption{{\bf {\sc Trace} Bovia Scenario: Episode 4 (Part 2).}
  On 9/9 08:35:35 the attacker runs the following programs: {\tt hostname},
  {\tt ifconfig} {\tt eth1.129}, {\tt ifconfig}, {\tt ifconfig} {\tt em1.192},
  {\tt ifconfig} {\tt em1.129}, and {\tt ls}. Later {\tt mkdir rootman} is run.
  The dropper program creates the file {\tt /home/steve/rootman/rmanager} and peforms {\tt chmod} on it. The attacker at this point runs the following programs using a remote shell: {\tt whoami}, {\tt hostname}, {\tt uname} {\tt -a}, {\tt ls} {\tt -laR} {\tt /home/steve} and the output is written to a file {\tt /home/steve/rootman/.dat}. Later {\tt cat} is run on it. The file is also opened with {\tt vi}.  The {\tt rmanager} program is then run with the {\tt .dat} file and the information is sent to 129.55.12.167:2525. Next the attacker checks the directory by running {\tt ls} and {\tt ls -la} and also checks running processes using
  {\tt ps} and {\tt ps} {\tt aux}. After that the attacker performs {\tt shred} on both the {\tt .dat} file and on {\tt rmanager}. The directory {\tt /home/steve/rootman} is then deleted. At 08:44:55 the attacker deletes the {\tt dropper} file.
}
\label{sribov42}
\end{figure}

The Bovia scenario covers a period of about 4 days, and
consists of four ``episodes.'' Figures~\ref{sribov1}--\ref{sribov3}
illustrate the first three episodes. The last episode is somewhat large,
and hence is broken up into two parts in Figures~\ref{sribov41} and
\ref{sribov42}.
\clearpage

\begin{figure}
\begin{center}
\vspace*{-4em}
\includegraphics[height=0.85\textheight]{{sbu-figs/sri_bovia1}}
\vspace*{-2em}
\end{center}
\caption{{\bf {\sc Trace} Bovia Scenario: Episode 1.} This episode starts with
    the download of the malicious file {\tt /home/steve/traffic\_gen/dropper}
    from 129.55.12.167:8000 and its execution on 9/6 at 17:58:47. Next the
    attacker gains a remote shell through 129.55.12.167:4430 and executes {\tt
      whoami}, {\tt hostname}, {\tt uname} {\tt -a}, {\tt ifconfig}, and {\tt netstat
      -nap}, and performs a {\tt cat} on {\tt /etc/hosts} and {\tt
      /etc/network/interfaces}. The dropper process creates {\tt /tmp/netrecon}
    and performs {\tt chmod} on it. Next, {\tt netrecon} is executed, which
    results in the creation of {\tt /tmp/netrecon.log} file. This file is read
    by the {\tt dropper} process and exfiltrated to 129.55.12.167:4430. Finally, at
    18:01:00, {\tt netrecon}, {\tt netrecon.log} and {\tt dropper} files are
    deleted from the system.
}
\label{sribov1}
\end{figure}

\begin{figure}
\begin{center}
\vspace*{-4em}
\includegraphics[width=0.95\textwidth]{{sbu-figs/sri_bovia2}}
\vspace*{-2em}
\end{center}
\caption{{\bf {\sc Trace} Bovia Scenario: Episode 2.} This episode starts with the download of {\tt /home/steve/traffic\_gen/dropper} from 129.55.12.167:8000 and its execution on 9/7 at 14:40:51. Next the attacker executes {\tt hostname}. The dropper process creates {\tt /tmp/netrecon} and {\tt /tmp/screengrab} files and performs {\tt chmod} on them. The {\tt netrecon} file is executed and it generates
  {\tt /tmp/netrecon.log} file. The execution of {\tt /tmp/screengrab}
  probably generates {\tt /tmp/sgout.png} but it was not seen in the data either
  the program failed  or the TA1 team failed to capture it. Next, {\tt netrecon.log} is read by the dropper process. After that the attacker deletes {\tt netrecon}, {\tt netrecon.log},
  and {\tt screengrab} and {\tt sgout.png} files. The attacker performs {\tt ls}, presumably to check if the files were deleted properly.  Finally, the {\tt  dropper} file is deleted at 14:48:48.
}
\label{sribov2}
\end{figure}

\subsubsection{Pandex Scenario}
The Pandex scenario covers a period of about 3 days. On 9/7 at 15:44:03, the
attacker first creates a directory {\tt /var/dropbear\_latest} and then downloads the
file {\tt dropbearLINUX.tar} to that directory using {\tt scp} from {\tt
  kududyn@128.55.12.167:/work/pandex\_software/dropbear/dropbearLINUX.tar}. The attacker performs {\tt tar xvf
  dropbearLINUX.tar}, which creates and {\tt chmod}'s the files {\tt
  dropbearkey}, {\tt dropbearscript} and {\tt dropbear}. Next, the attacker
creates {\tt /etc/dropbear} directory and executes {\tt dropbearkey} program,
which generates {\tt dropbear\_rsa\_host\_key}, {\tt dropbear\_ecdsa\_host\_key}
and {\tt dropbear\_dss\_host\_key} in that directory. The attacker then runs  {\tt ./dropbear} {\tt -p} {\tt 1337}, which uses all the three
keys generated by {\tt dropbearkey} program and performs exfiltration. On 9/9 at
10:23:24, the attacker performs a cleanup by running the {\tt shred} program on
the three keys and then performing {\tt rm} {\tt -rf} on them. In addition, {\tt
  rm} {\tt -rf} is done on the {\tt /etc/dropbear} folder and the {\tt
  /var/dropbear\_latest} folder.

\begin{figure}
\begin{center}
\vspace*{-2em}
\includegraphics[width=1.01\textwidth]{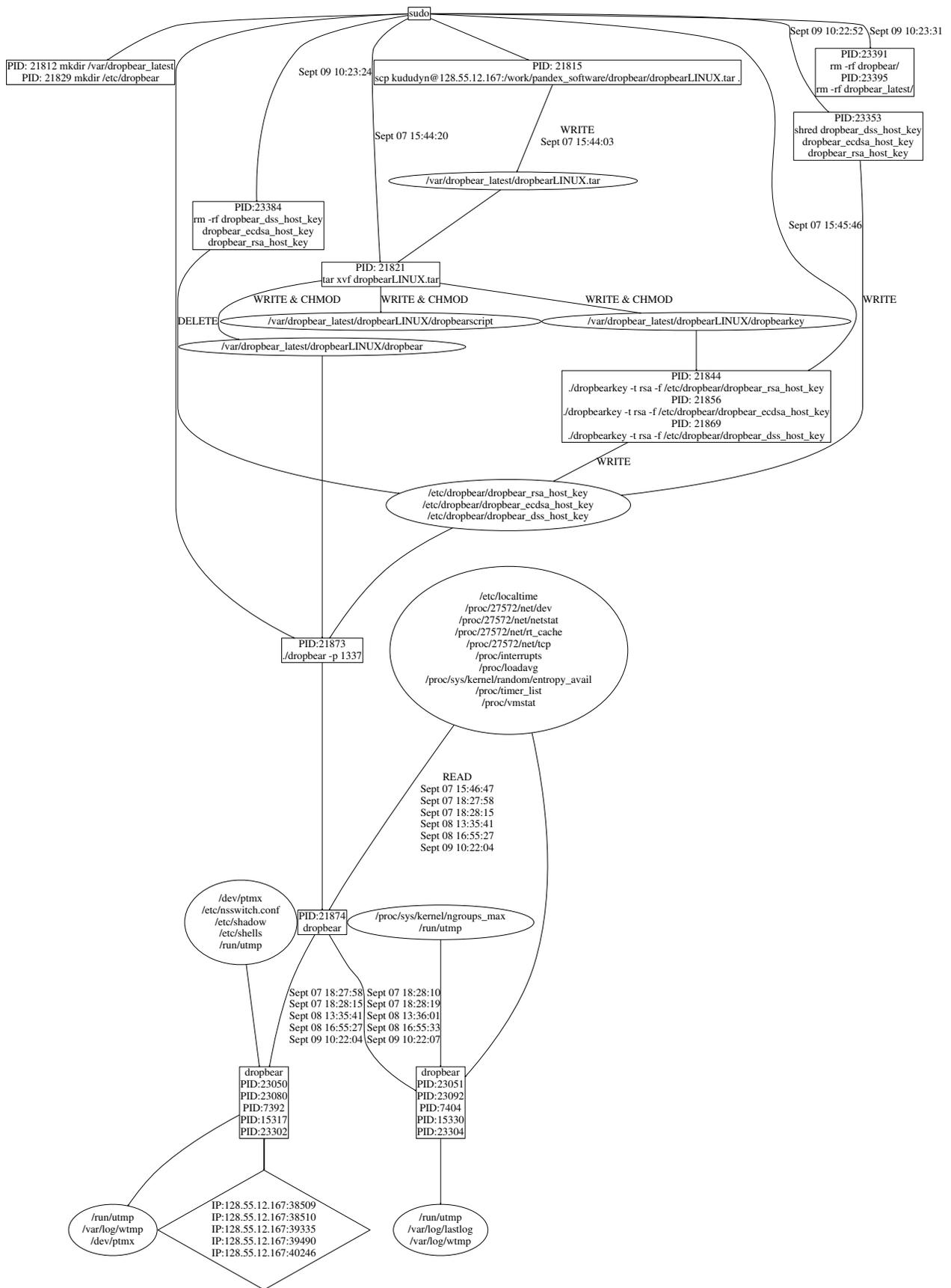}
\vspace*{-3em}
\captionsetup{singlelinecheck=off}
\caption{{\bf {\sc Trace} Pandex Scenario.} 
}
\label{sripandex}
\end{center}
\end{figure}

The attack consist of several information gathering sessions, followed
by exfiltration. These exfiltrations occur to the following IP addresses
at the following times:
\begin{itemize}
     \item IP:128.55.12.167:38509  on 9/7 18:27:58
     \item IP:128.55.12.167:38510  on 9/7 18:28:15
     \item IP:128.55.12.167:39335  on 9/8 13:35:41
     \item IP:128.55.12.167:39490  on 9/8 16:55:27
     \item IP:128.55.12.167:40246  on 9/8 10:22:04
\end{itemize}
The Figure \ref{sripandex} describes the attack.
\pagebreak

\subsubsection{Stretch Scenario}
\begin{figure}[h]
\begin{center}
\vspace*{-4em}
\includegraphics[width=0.75\textwidth]{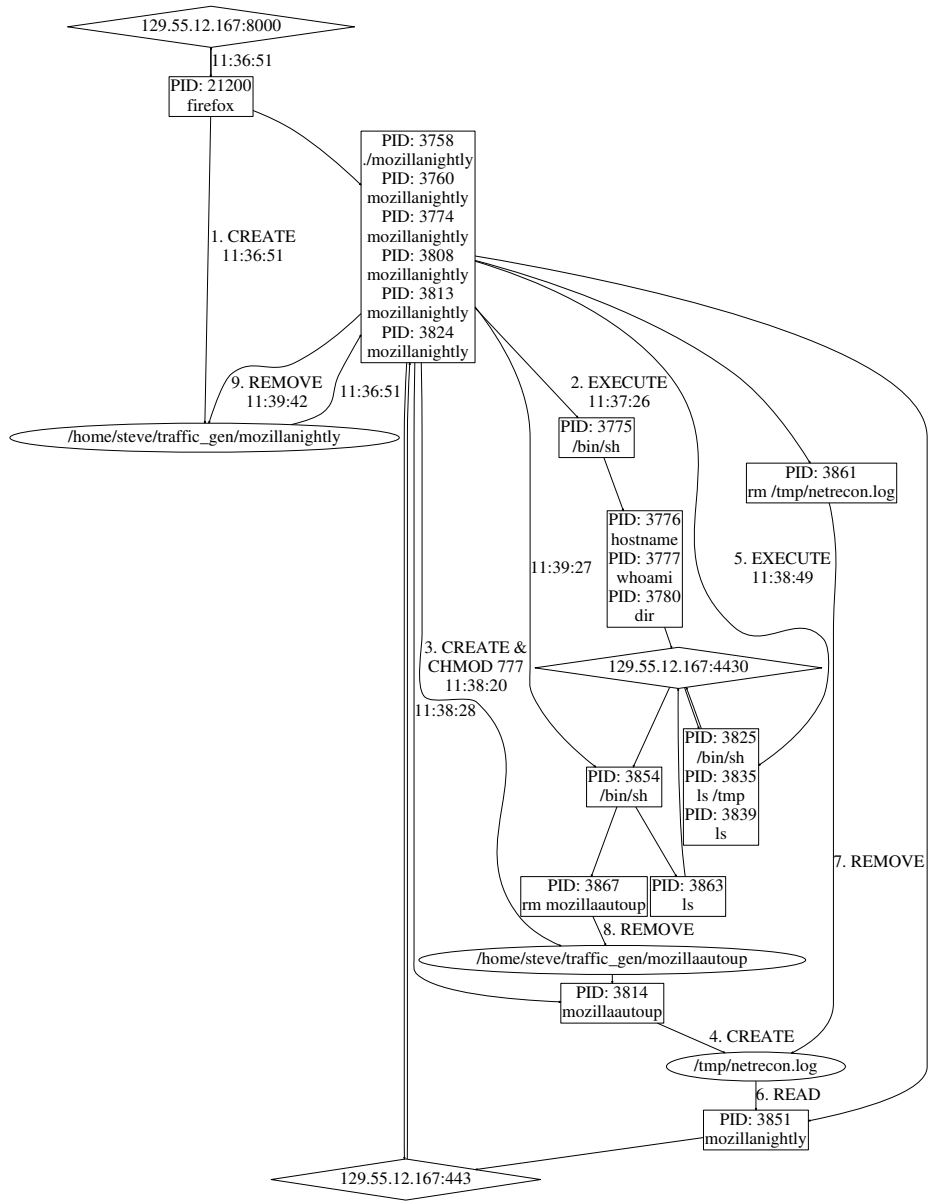}
\vspace*{-2em}
\end{center}
\caption{{\bf {\sc Trace} Stretch Scenario: Episode 1.} The attack begins with
  what seems to be a browser compromise of Firefox at 11:36:51, which causes a
  malicious executable {\tt mozillanightly} to be downloaded from IP address
  129.55.12.167:8000 and executed. This allows the attacker to gain a remote
  shell through IP address 129.55.12.167:4430. The attacker executes the
  programs {\tt hostname} and {\tt whoami}. At 11:38:20, {\tt mozillanightly}
  creates {\tt /home/steve/traffic\_gen/mozillaautoup} and performs a {\tt
    chmod} on it with permission 777. This file is executed at 11:38:28 by {\tt
    mozillanightly} and a file {\tt /tmp/netrecon.log} is created and written
  to. The attacker performs {\tt ls} and {\tt ls} {\tt /tmp}, probably to check through the remote shell
  whether the files were created. The file {\tt
    netrecon.log} is read, and the information is sent to the IP address
  129.55.12.167:443. At 11:39:27, the attacker deletes {\tt
    /tmp/netrecon.log} and performs {\tt ls} again through the remote
  shell. At 11:39:36 the file {\tt /home/steve/traffic\_gen/mozillaautoup} is
  deleted. At 11:39:42 the malicious executable {\tt
    /home/steve/traffic\_gen/mozillanightly} is deleted. }
\label{sristr_1}
\end{figure}

\begin{figure}[h]
\begin{center}
\vspace*{-4em}
\includegraphics[width=0.9\textwidth]{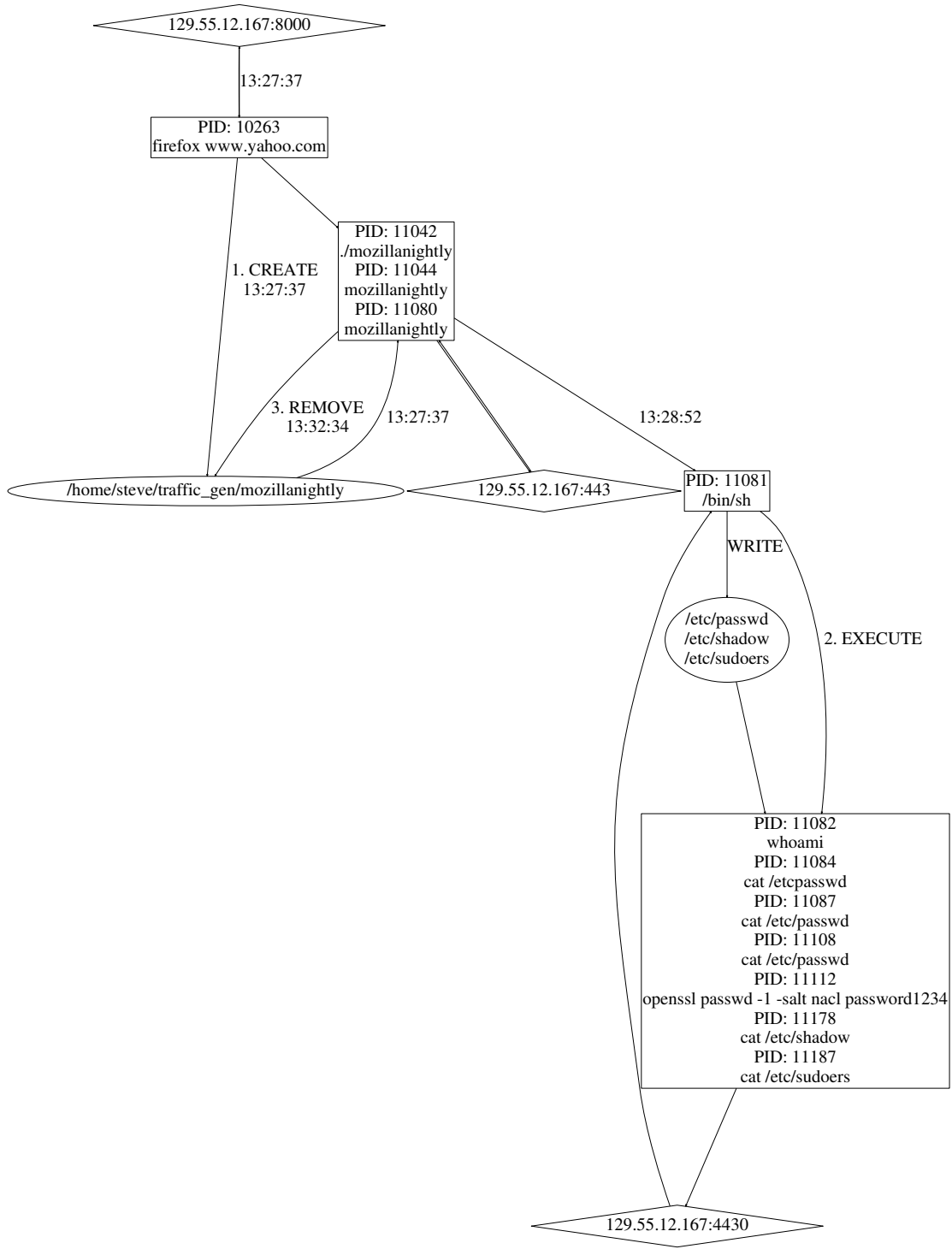}
\vspace*{-2em}
\end{center}
\caption{{\bf {\sc Trace} Stretch Scenario: Episode 2.} In the second attack,
  compromised firefox downloads the malicious executable {\tt
    /home/steve/traffic\_gen/mozillanightly} again at 13:27:37 through IP
  address 129.55.12.167:8000 and executes it. This allows the attacker to gain a
  remote shell again through IP address 129.55.12.167:4430. The attacker
  first executes {\tt whoami} program. Then the attacker performs {\tt cat}
  on the {\tt /etc/passwd} file and writes to it. Next the {\tt cat} program is
  run on the {\tt /etc/shadow} file and also written to. After that, the attacker writes
  to the {\tt /etc/sudoers} file at 13:32:18 and executes {\tt cat} on the file.
  The malicious downloaded file {\tt /home/steve/traffic\_gen/mozillanightly} is
  deleted at 13:32:34. }
\label{sristr_2}
\end{figure}

The stretch scenario consists of two ``episodes.'' Figures \ref{sristr_1} and
\ref{sristr_2} illustrates these episodes. 

\end{document}